\documentclass[aps,prd,showpacs,amsmath,showkeys,twocolumn,floatfix,amssymb, preprintnumbers, nofootinbib, superscriptaddress]{revtex4} 
\usepackage{epsfig,dcolumn}
\usepackage{graphicx}
\usepackage{comment} 
\usepackage{verbatim}
\usepackage[usenames]{color}
\usepackage{bm}
 \usepackage{ifpdf}

\begin{document}

\title{ Propagation of particles on a torus }

\author{Peng~Guo}
\email{pguo@csub.edu}

\affiliation{Department of Physics, California State University, Bakersfield, CA 93311, USA}
\affiliation{Kavli Institute for Theoretical Physics, University of California, Santa Barbara, CA 93106, USA}

\date{\today}

\begin{abstract} 
In this study, based on the variational principle and Faddeev method, we present a general framework for finding the propagating solutions of multiple interacting particles on a torus. Two different versions of multiparticle secular equations are presented. Version one shows  how the propagating solutions on a torus and the infinite volume dynamics are connected. The second version may be more suitable and robust for the task of lattice quantum chromodynamics data analysis. The proposed formalism may also be useful for studying the effects of few-body interactions on the electronic band structure in condensed matter physics. 
   \end{abstract}


\maketitle


{\bf Introduction.}--Few-body interactions are crucial elements in many areas of modern physics, e.g., extracting the mass ratio of light quarks from experimental data requires the precise determination of three-body dynamics \cite{Kambor:1995yc,Anisovich:1996tx,Schneider:2010hs,Guo:2015zqa,Guo:2016wsi,Colangelo:2016jmc}. Few-body systems may also exhibit distinct features and unexpected long-range few-body effects, such as the Effimov effect \cite{Efimov:1970zz} and nuclear halo effect \cite{Zhukov:1993aw}. In recent years, tremendous progresses on the studies  of the interactions of multiple hadron particles based on first principles calculation using lattice quantum chromodynamics (QCD) have been made in nuclear/hadron physics. However, extracting multiparticle dynamics from lattice simulations is  still mostly limited to the two-particle sector, and the two-body finite volume formalism is largely based on the L\"uscher formula and its variants \cite{Luscher:1990ux,Rummukainen:1995vs,Christ:2005gi,Bernard:2007cm,Bernard:2008ax,He:2005ey,Lage:2009zv,Doring:2011vk,Briceno:2012yi,Hansen:2012tf,Guo:2012hv,Guo:2013vsa,Kim:2005gf,Briceno:2014uqa,Romero-Lopez:2018zyy}. Three-body problems have been addressed using various approaches \cite{Kreuzer:2008bi,Kreuzer:2009jp,Kreuzer:2012sr,Polejaeva:2012ut,Briceno:2012rv,Hansen:2014eka,Hansen:2015zga,Hansen:2016fzj,Hammer:2017uqm,Hammer:2017kms,Meissner:2014dea,Briceno:2017tce,Sharpe:2017jej,Mai:2017bge,Romero-Lopez:2018rcb,Guo:2016fgl,Guo:2017ism,Guo:2017crd,Guo:2018ibd,Guo:2018xbv,Mai:2018djl,Briceno:2018aml,Romero-Lopez:2019qrt,Blanton:2019vdk,Doring:2018xxx,Pang:2019dfe}, where most of these developments are along the line of  building connections between infinite volume scattering amplitudes and long-range correlations due to the periodic structure of the cubic lattice. Formulating the quantization conditions using infinite volume scattering amplitudes as the inputs present a more conventional foundation, but both the infinite and finite-volume dynamics must be handled  simultaneously, which may   be computationally challenging.

In this letter, we present a framework with the potential to overcome some technical obstacles and provide a more general and convenient method for analyzing lattice QCD simulations of multiparticle dynamics. The framework presented in this letter is based on the original variational approach proposed by \cite{Guo:2018ibd}, which employs a multiparticle Schr\"odinger equation in a differential equation representation. One of the key components of finite volume physics is that a periodic boundary condition must be satisfied by the solutions. In the method proposed by \cite{Guo:2018ibd}, the periodicity of the solutions is achieved by constructing variational basis functions based on the linear superposition of infinite-volume solutions centered at each image of the cubic box, $\phi_J (x) = \sum_{n\in \mathbb{Z}} \psi_J (x+ nL) $, where $\phi_J (x)$ represents the finite volume variational basis, $\psi_J (x)$ is one of the independent infinite volume solutions labeled with a quantum number $J$, and the symbol $L$ denotes the size of the periodic box. The desired boundary condition is now satisfied by all $\phi_J (x)$, but these $\phi_J (x)$ are typically not solutions of the finite volume multiparticle Schr\"odinger equation. Therefore, the variational principle must be applied and the variational trial function may be given by $\phi (x) = \sum_J c_J \phi_J (x)$ (see \cite{Guo:2018ibd}).  
 Hence, finite volume and infinite volume physics are linked together by constructing finite volume basis functions from infinite volume solutions. Thus, the original idea is similar to other previously proposed methods \cite{Kreuzer:2008bi,Kreuzer:2009jp,Kreuzer:2012sr,Polejaeva:2012ut,Briceno:2012rv,Hansen:2014eka,Hansen:2015zga,Hansen:2016fzj,Hammer:2017uqm,Hammer:2017kms,Meissner:2014dea,Briceno:2017tce,Sharpe:2017jej,Mai:2017bge,Romero-Lopez:2018rcb,Guo:2016fgl,Guo:2017ism,Guo:2017crd,Guo:2018ibd,Guo:2018xbv,Mai:2018djl,Briceno:2018aml,Romero-Lopez:2019qrt,Blanton:2019vdk,Doring:2018xxx,Pang:2019dfe} where infinite volume solutions are used as inputs for finite volume physics. Finding infinite volume solutions alone can be cumbersome. In this letter, we show that an integral equation representation of the variational approach may be more favorable for finding solutions of finite volume multiparticle dynamics. The periodic boundary conditions are automatically satisfied by the solutions of the Lippmann--Schwinger type equation. Hence, the key step in our original approach involving the construction of finite volume basis functions means that imposing boundary conditions is no longer required. Ultimately, the quantization condition may be given in a form that requires no specific choice of the variational basis and it depends only on the interaction potentials and finite volume Green's functions. In order to conduct lattice data analysis, the interaction potentials may be treated as inputs and the discrete energy spectrum of multiparticle interactions in a finite volume may be found without knowing the infinite volume scattering amplitudes or wave functions. After extracting the interactions from the lattice results, infinite volume dynamical quantities such as the scattering amplitudes may be computed separately.

The key to studying multiparticle dynamics in lattice QCD simulations is finding the propagating solutions for multiple interacting particles in a periodic box, which is similar to studying electronic band structures in condensed matter physics. Therefore, the proposed framework may also be useful for studying few-body effects on electronic band structures. 
In addition, the formalism presented in this letter may provide a natural framework for including coupled-channel effects in different particle sectors. The key concept employed in the integral equation representation of the variational approach is illustrated with a simple example of a single particle propagating through a periodic potential.

{\bf Single particle propagation on a one dimensional torus, L\"uscher formula, and variational approach.}-- We start our discussion with a simple quantum mechanical problem involving two particles moving on a torus. By factorizing out the center of mass motion, the problem of two propagating particles on a torus may be mapped onto a problem of a single particle tunneling through a periodic potential. It is sufficient to use this basic example to illustrate the connection with the L\"uscher formula and the key ideas in the variational approach in a pedagogical  manner without excessive technical details.

{\it Single particle tunneling though a periodic potential: }
The problem of a quantum mechanical particle propagating in a periodic box with size $L$ and interacting with a short-range spatially symmetric potential, $V(-x) = V(x)$, is described by the Schr\"odinger equation:
\begin{equation}
\left ( E +\frac{ \nabla^2 }{2 m}  \right )  \phi(x,E) = V(x) \phi(x,E) , \ \  x \in [ -\frac{L}{2}, \frac{L}{2}], \label{schrodingerfv}
\end{equation}
where the wave function is required to satisfy the periodic boundary condition:
\begin{equation}
\phi(  - \frac{ L}{2},E)  = e^{-i Q L} \phi( \frac{L}{2},E)  , \ \  \phi'(  - \frac{ L}{2},E)  = e^{-i Q L} \phi'( \frac{L}{2},E)   .   \label{boundary1D}
\end{equation}
 $Q$ may be related to the CM momentum of two particles, $P= \frac{2\pi d}{L}$ $ (d \in \mathbb{Z})$, by $Q= \frac{P}{2}$ in lattice QCD computations \cite{Guo:2012hv} or the quasi-momentum of a particle in condensed matter physics \cite{Ziman:1964}. For general discussion purposes, in this study, we assume that $Q$ is an arbitrary real number. A conventional approach for solving the problem described above involves expressing the finite volume wave function as the linear superposition of all the independent solutions of the corresponding Schr\"odinger equation in infinite space: $\phi(x,E)   = \sum_J c_J \psi_J(x,E)  $, where $\psi_J$ is the solution of the following equation:
\begin{equation}
\left ( E +\frac{ \nabla^2 }{2 m}  \right )  \psi_J (x,E) = V(x) \psi_J(x,E) ,   \ \ x \in [ - \infty,  \infty]
\end{equation}
 corresponding to a free incoming wave $\psi_J^{(0)}$. Each single $\psi_J$ usually does not satisfy the periodic boundary conditions in Eq. (\ref{boundary1D}), and thus the solution of Eq. (\ref{schrodingerfv}), $\phi=\sum_J c_J \psi_J$, must be obtained by imposing periodic boundary conditions and finding the correct coefficients, $c_J$, to satisfy the boundary condition for the wave function $\phi$. Therefore, imposing the boundary condition ultimately yields the quantization condition that determines the allowed energy spectrum and the solutions of the finite volume wave function in terms of infinite volume wave solutions. In one dimension, only two independent solutions are present in the infinite volume \cite{Guo:2013vsa,Guo:2016fgl}, $\psi_{\pm} ( - x ,E) = \pm \psi_{\pm} (x,E)$, which have the asymptotic form:
 \begin{equation}
 \psi_{\pm} (x,E) \stackrel{|x| \sim \frac{L}{2}  }{ \rightarrow } \psi^{(0)}_{\pm} (x,E) + i t_{\pm} (k) e^{i k |x |} Y_{\pm} (x), \  k =\sqrt{2 m E}, \label{psiasym}
 \end{equation}
 where $\psi^{(0)}_{+} (x,E) = \cos k x$, $\psi^{(0)}_{-} (x,E) = i \sin k x$, $Y_+(x) =1$ and $Y_-(x) = \frac{x}{|x|}$. The physical scattering amplitudes, $t_\pm (k)$, are defined by  
$ t_\pm(k) =- \frac{1}{2k} \int_{-\infty}^{\infty} d x e^{- i k x} V(x) \psi_{\pm} (x,E)$. Thus, by using the asymptotic form of $\psi_{\pm}$ given in Eq. (\ref{psiasym}), the quantization condition can be found easily after the boundary conditions are imposed: 
\begin{align}
  &  \tan^2 \frac{QL}{2} \left [ \frac{1+ e^{- i k L}}{2 i } +  t_+(k) \right ]   \left [ \frac{1+ e^{- i k L} }{2 i}+   t_-(k) \right ]  \nonumber \\
 & +   \left [ \frac{ 1- e^{- i k L}}{2 i } +   t_+(k) \right ]   \left [ \frac{1- e^{- i k L}}{2 i } +   t_-(k) \right ] = 0 . \label{quantization1D}
\end{align}
In addition, the ratio of $c_+$ and $c_-$ is obtained as:
\begin{equation}
\frac{c_+}{c_-} =  - i  \cot \frac{QL}{2}  \frac{  1- e^{- i k L} + 2 i t_-(k)   }{   1+ e^{- i k L}+2 i  t_+(k)    } . \label{coefratio}
\end{equation}
Hence, the problem of a single particle propagating though a periodic potential is now considered to be solved completely because both the quantization condition and coefficients of linear superposition are obtained. The quantization condition in Eq. (\ref{quantization1D}) may be regarded as a generalized L\"uscher formula in one dimension \cite{Guo:2013vsa,Guo:2016fgl}. In general, for an arbitrary value of $Q$, both solutions of $\psi_{\pm}$ must contribute to ensure that $\phi = \sum_{J=\pm} c_J \psi_J$ satisfies the periodic boundary condition. To illustrate this point further, let us also consider two particles interacting with a $\delta$-function potential in three dimensions as another example \cite{Guo:2018ibd}. With a $\delta$-function potential, only the $S$-wave scattering amplitude contributes and the secular equation is simply given by $ c_J \left [ i + \frac{1}{t_0 (k)}   \right ]= c_0  \mathcal{M}^{(\mathbf{ Q})}_{[J], [0]} (k) $, where $ \mathcal{M}^{(\mathbf{ Q})}_{[J],[J']} $ denote the partial wave expansion coefficients of the finite volume Green's function (see \cite{Guo:2012hv,Guo:2018ibd}). In a finite volume, all the partial waves must be summed up with the correct ratios to satisfy the periodic boundary condition and the final finite volume wave function is proportional to the periodic finite volume Green's function: $\phi(\mathbf{ x},E) \propto t_0 (k) G_L^{(Q)} (\mathbf{ x}, E)$.

{\it Finite volume Lippmann--Schwinger equation: }
In one dimension, the conventional boundary condition matching procedure presents a clear demonstration for the quantum mechanical periodic boundary considered in the example above, but the matching boundary condition procedure is more technically tedious in higher dimensions and multiparticle sectors. Hence, the differential equation, Eq. (\ref{schrodingerfv}), and boundary conditions, Eq. (\ref{boundary1D}),  together may   be replaced by a finite volume Lippmann--Schwinger type equation:
\begin{equation}
 \phi(x,E) =  \int_{-\frac{L}{2}}^{\frac{L}{2}} d x' G^{(Q)}_L (x-x', E) V(x') \phi(x',E), \label{homoeq}
\end{equation} 
 where the finite volume Green's function is given by:
 \begin{equation}
 G^{(Q)}_L (x,E) = \frac{1}{L} \sum_{n \in \mathbb{Z}}^{p = \frac{2\pi n}{L} +Q} \frac{e^{i p x}}{E - \frac{p^2}{2m}} .
 \end{equation}
The periodic boundary conditions for $\phi$ in Eq. (\ref{boundary1D}) are automatically satisfied due to the periodicity of $G^{(Q)}_L $:
  \begin{equation}
 G^{(Q)}_L (x + L,E) = e^{i QL} G^{(Q)}_L (x ,E).
 \end{equation}
 Again, by using the fact that $\phi  = \sum_\alpha c_J \psi_J   $ because the solutions of $\psi_{\pm}$ are given, then solving the finite volume Lippmann--Schwinger equation can be transformed into the problem of finding the allowed energy values and $c_{\pm}$ that satisfy the following equation:
 \begin{align}
 & \sum_{J=\pm} c_J   \psi_J(x,E)   \nonumber \\
 & =   \sum_{J=\pm} c_J   \int_{-\frac{L}{2}}^{\frac{L}{2}} d x' G^{(Q)}_L (x-x', E) V(x') \psi_J(x',E) . \label{LSfv}
 \end{align} 
 To obtained the L\"uscher formula again from Eq. (\ref{LSfv}), we may use the asymptotic form of $\psi_J$ in Eq. (\ref{psiasym}) and the analytic expression of the finite volume Green's function:
  \begin{align}
&  G^{(Q)}_L (x,E)  \nonumber \\
& = - \frac{ 2 m i}{2 k} \left  [ e^{i k |x|} + \frac{e^{i k x}  }{e^{i(Q- k)L} -1} + \frac{e^{-i k x}  }{e^{ -i(Q+k)L} -1} \right ].
 \end{align}
 Thus, we find that:
 \begin{align}
 &  c_+ \cos k x + c_- i \sin k x \nonumber \\
 & = \frac{e^{ i k x}}{e^{i (Q-k) L}-1} \left [ c_+ i t_+ (k) + c_- it_-(k)\right ]  \nonumber \\
 &+   \frac{  e^{ - i k x}}{e^{ -i (Q+k) L}-1}\left [ c_+ i t_+ (k) - c_- it_-(k)\right ]. \label{luscher1D}
 \end{align}
 $\cos kx$ and $ \sin kx$ are two independent bases, so their coefficients in Eq. (\ref{luscher1D}) must match on both sides of the equation, thereby ultimately yielding the quantization condition and the ratio of $c_+/c_-$ given in Eq. (\ref{quantization1D}) and Eq. (\ref{coefratio}), respectively.

{\it Variational approach in integral equation representation: }
The L\"uscher formula provides an explicit connection between the periodic structure of the lattice and the physical scattering amplitude of the infinite volume. As illustrated in the previous discussion, obtaining these explicit relations requires knowledge of the asymptotic form of the multiparticle wave function and the asymptotic properties of the finite volume Green's function. Unfortunately,  finding the asymptotic forms of the wave function and Green's function is not simple for multiparticle interactions. In particular, pairwise interactions behave as long-range interactions in multiparticle sectors, thereby making the study of asymptotic behaviors even more difficult. The original variational approach in differential equation representation \cite{Guo:2018ibd} was developed to provide a numerical framework for finding solutions to the multiparticle dynamics in a finite volume by dispensing with the analytical form of the L\"uscher formula. Thus, in order to achieve the same goal, in this study, the variational principle is applied to the finite volume Lippmann-Schwinger equation in Eq. (\ref{LSfv}), and $\phi  = \sum_\alpha c_J \psi_J   $ may be regarded as a variational trial function. By integrating both sides of Eq. (\ref{LSfv}) by $ \int_{-\frac{L}{2}}^{\frac{L}{2}}  d x  \psi^*_{J'}(x,E) $, a secular equation in a periodic box is obtained:
\begin{equation}
\sum_J \left [ S_{ J',J} (E) - H^{(Q)}_{J', J} (E) \right ] c_J =0, \label{seq}
\end{equation}
where $S_{ J',J} (E) =  \int_{-\frac{L}{2}}^{\frac{L}{2}}  d x   \psi^*_{J'}(x,E)  \psi_J(x,E) $, and: 
 \begin{align}
 & H^{(Q)}_{J',J}   =     \int_{-\frac{L}{2}}^{\frac{L}{2}}  d x  d x'   \nonumber \\
 & \quad \quad  \times   \psi^*_{J'} (x,E) G^{(Q)}_L (x-x', E) V(x') \psi_J(x',E)   .  
 \end{align} 
 Therefore, in the variational approach, the quantization condition is given by:
 \begin{equation}
 \det  \left [ S_{ J',J} (E)- H^{(Q)}_{J',J} (E) \right ] =0, \label{vardet}
 \end{equation}
  and the ratio of the coefficients can also be found as: 
  \begin{equation}
  \frac{c_+}{c-} = \frac{  S_{ +,-} (E)- H^{(Q)}_{+,-} (E)   }{  S_{ +,+} (E)- H^{(Q)}_{+,+} (E)    },
  \end{equation}
    which are equivalent to  Eq. (\ref{quantization1D}) and Eq. (\ref{coefratio}), respectively. 
 
 The variational approach has the advantage that no explicit asymptotic expressions are required for the wave functions and finite volume Green's function, and thus it may provide a convenient numerical approach to finite volume problems. However, this approach is still computationally intensive in practice. The infinite volume solutions must be found first and then used as inputs to compute all of the matrix elements in Eq. (\ref{vardet}). Therefore, this approach is still highly numerically challenging and it might not be the most suitable form for the purpose of lattice data fitting.

The problem in Eq. (\ref{vardet}) are all caused by the selection of a specific form of variational trial function: $\phi  = \sum_\alpha c_J \psi_J   $, and thus Eq. (\ref{vardet}) depends on the infinite volume solutions. 
 If we only need to obtain the energy spectrum, the quantization condition may be arranged in a form that is free of a specific form of the trial basis, which depends only on the interaction potentials and finite volume Green's function. Thus, the quantization condition given in this manner may be more suitable for the purpose of lattice data fitting. To illustrate this idea, 
  the finite volume Lippmann--Schwinger equation, Eq. (\ref{coefratio}), is first transformed into a homogeneous matrix equation by discretizing it in a finite box:
 \begin{equation}
      \sum_j  \left [ \delta_{i,j} - G^{(Q)}_L (x_i-x_j, E) V(x_j) \right ]  \phi_j =0,  \label{LSeqfvdiscrete1D}
      \end{equation} 
where $x_j = a j \in [-\frac{L}{2}, \frac{L}{2}]$ denote the discrete positions in the finite volume, $a$ may be regarded as the finite lattice spacing, and $\phi_j = \phi(x_j,E)$.
In this case, Eq. (\ref{LSeqfvdiscrete1D}) may be regarded as secular equations with coefficients of $\phi_j$ in the same manner as the secular equation given in Eq. (\ref{seq}), and thus the quantization condition is simply given in terms of the potential and finite volume Green's function by:
\begin{equation}
\det  \left [ \delta_{i,j} - G^{(Q)}_L (x_i-x_j, E) V(x_j) \right ] =0. \label{qcGV}
\end{equation} 
The normalized finite volume wave function, $\phi$, may also be solved numerically. By treating the potentials as inputs instead of infinite volume wave functions, the quantization condition given in Eq. (\ref{qcGV}) does not require any extra effort to find solutions for the infinite volume amplitudes or wave functions, and thus it may be more efficient for lattice QCD data analysis in practice.

{\bf  Multiparticle propagation on a torus and variational approach.}-- Using finite volume integral equations, such as the Lippmann--Schwinger equation, in a higher dimensional space or for multiparticle dynamics has clear advantages for implementing periodic boundary conditions. The general form for a multiparticle interaction in a finite periodic box may be given by an integral equation defined in a single cell of a periodic box:
\begin{equation}
 \Phi    = G_L V   \Phi    , \label{integeq}
\end{equation}
where the multiparticle finite volume Green's function, $G_L$, and finite volume wave function, $\Phi$, share the same periodic boundary condition. By applying the same methods described above, the finite volume wave function may be given as the sum of all possible independent propagating solutions in an infinite volume, $\Phi = \sum_J c_J \Psi_J$. $\Psi_J$ satisfies the integral equation:
\begin{equation}
 \Psi_J     =  \Psi^{(0)}_J   +  G_0 V   \Psi_J    , \label{freeintegeq}
\end{equation}
where $G_0$ denotes the infinite volume Green's function and $\Psi^{(0)}_J$ represents an incoming wave.
We need to transform the integral equation, Eq. (\ref{integeq}), into a secular equation:
\begin{equation}
\sum_J \langle \Psi_{J'} | \mathbb{I} - G_L V  | \Psi_J \rangle  c_J =0, \label{integseq}
\end{equation}
where $  \Psi_J     = \left ( \mathbb{I}  -  G_0 V\right )^{-1}   \Psi^{(0)}_J  $. Both the quantization condition and solutions of $\Phi$ may be obtained from the secular equation, Eq. (\ref{integseq}). Typically, the number of independent solutions is infinite in a higher dimensional space or multiparticle sectors because the only constraint on the kinematics is the total energy of the particles. For example, single a particle in three dimensions, the incoming waves can be selected as partial waves, $\Psi^{(0)}_{J= [J]} (\mathbf{ r})  = j_J(k r) Y_{[J]} (\mathbf{ r})$. Hence, the variational basis must be truncated in practical computations.

{\it Faddeev method and multiparticle propagating solutions in an infinite volume: } 
In terms of the multiparticle dynamics, especially when pairwise interactions are involved, the disconnected diagrams lead to non-compactness for the integral kernel in Eq. (\ref{freeintegeq}) and direct difficulty solving Eq. (\ref{freeintegeq}). Faddeev procedures \cite{Faddeev:1960su,Faddeev:1965} are usually employed to transform a single integral equation, Eq. (\ref{freeintegeq}), into a set of coupled equations with a well-defined kernel. The variational approach and Faddeev method described in the following are not limited to three-body problems, but in order to simplify the presentation, we constrain our discussion to only the three-body problem with both pairwise interactions and three-body short-range interactions. In an infinite volume, the scattering solutions for three particles are given by:
\begin{equation}
 \Psi_J     =  \Psi^{(0)}_J   +  G_0 (v_1 + v_2 + v_3 + v_4 )   \Psi_J    ,  \label{free3bintegeq} 
\end{equation}
where $v_{\alpha=1,2,3} $ denote pairwise interactions between the pairs $(\beta \gamma)$ and $\alpha \neq \beta \neq \gamma $, and $v_4$ represents the three-body short-range interactions involving all three particles. Thus, by introducing the scattering amplitudes, $ T_{\alpha,J} = - v_\alpha \Psi_J$, and the total scattering amplitude, $T_J = \sum_{\alpha=1}^4  T_{\alpha,J}  = - V  \Psi_J  $, the three-body Lippmann--Schwinger equation, Eq. (\ref{free3bintegeq}), is transformed into a set of coupled equations:
\begin{equation}
     T_{\alpha, J}      = - v_\alpha    \Psi^{(0)}_J    + v_\alpha G_0  \sum_\beta   T_{\beta, J}     .
\end{equation}
Defining operators, 
\begin{equation}
t_{\alpha} =  - (\mathbb{I} - v_\alpha G_0)^{-1} v_\alpha,
\end{equation}
and the Faddeev equations are obtained:
\begin{equation}
     T_{\alpha,J}     + t_\alpha G_0   \sum_{\beta \neq \alpha}  T_{\beta, J}       = t_\alpha   \Psi^{(0)}_J    . \label{faddeev}
\end{equation}
The Faddeev equations in Eq. (\ref{faddeev}) are standard Fredholm-type equations with a well-defined kernel and they can be solved with standard procedures \cite{Faddeev:1960su,Faddeev:1965,Gloeckle:1983,Faddeev:1993,Gloeckle:1995jg}.

{\it  Multiparticle secular equations in a finite volume: } 
The Faddeev procedures may also be applied in a finite volume in a similar manner, where we may introduce the finite volume amplitudes as $ T^{(L)}_{\alpha} = - v_\alpha \Phi $, and the total finite volume $T$-amplitude as $T^{(L)} = \sum_{\alpha=1}^4 T^{(L)}_{\alpha}  = - V \Phi$. Therefore, Eq. (\ref{integeq}) is transformed into:
\begin{equation}
     T_{\alpha}^{(L)}      =  v_\alpha G_L   \sum_\beta T^{(L)}_\beta     . \label{Faddeevfv}
\end{equation}
If we also introduce the finite volume operators as: 
\begin{equation}
t^{(L)}_{\alpha} =  - (\mathbb{I} - v_\alpha G_L)^{-1} v_\alpha,
\end{equation}
then  a finite volume version of the Faddeev equations is given by:
\begin{equation}
     T^{(L)}_{\alpha}     + t^{(L)}_\alpha G_L   \sum_{\beta \neq \alpha}  T^{(L)}_{\beta}       = 0   .  \label{Faddeevfvscattamp}
\end{equation}
Thus, in a finite volume, by applying the assumption that $\Phi=\sum_J c_J \Psi_J$, or equivalently that $T^{(L)}_{\alpha}   = \sum_J c_J T_{\alpha,J}$, the multiparticle secular equations are obtained as:
\begin{equation}
\sum_J  \left [  \sum_{\alpha, \beta}  \langle T_{\alpha, J'}  | \delta_{\alpha, \beta } (\mathbb{I} -  t^{(L)}_\alpha G_L ) +  t^{(L)}_\alpha G_L | T_{\beta, J} \rangle  \right ] c_J=0, \label{multibodyseq}
\end{equation}
where $   T_{\alpha,J}    = \left [ \delta_{\alpha,\beta} (\mathbb{I} - t_\alpha G_0) + t_\alpha G_0 \right ]^{-1} t_\alpha   \Psi^{(0)}_J  $. Therefore, the energy spectra of multiple particles propagating on a torus are determined by the quantization condition: 
\begin{equation}
\det \left [  \sum_{\alpha, \beta}  \langle T_{\alpha, J'}  | \delta_{\alpha, \beta } (\mathbb{I} -  t^{(L)}_\alpha G_L ) +  t^{(L)}_\alpha G_L | T_{\beta, J} \rangle  \right ] =0, \label{multibodyqc1}
\end{equation}
or in terms of the potentials instead of $t^{(L)}_\alpha$ operators as:
\begin{equation}
\det \left [  \sum_{\alpha, \beta}  \langle T_{\alpha, J'}  | \delta_{\alpha, \beta }  - v_\alpha G_L | T_{\beta, J} \rangle  \right ] =0.   \label{multibodyqc2}
\end{equation}
The finite volume wave functions are also solved by secular equations, Eq. (\ref{multibodyseq}).

The secular equations in Eq. (\ref{multibodyseq}) may provide a sound numerical framework for finding propagating solutions for multiparticles on a torus, but this method is still computationally intensive and may not be highly efficient for data fitting in practice. Thus, for practical purposes, it is probably more effective to directly solve the finite volume Faddeev equations in Eq. (\ref{Faddeevfv}) or Eq. (\ref{Faddeevfvscattamp}) by either discretizing in the finite coordinate space or by Fourier transformation into a finite volume momentum space with discrete free lattice momenta. For instance, Eq. (\ref{Faddeevfv}) may be transformed into a homogeneous matrix equation:
\begin{equation}
  \sum_{\beta, j}  \left [ \delta_{\alpha, \beta}  \delta_{i,j}      -  ( v_\alpha G_L   )_{i, j}  \right ] T^{(L)}_{\beta, j}=0     ,  \label{Faddeevfvdiscret}
\end{equation}
where the index $(i,j)$ denotes the discretized grid positions in a finite box. Therefore, the discretized finite volume Faddeev equation in Eq. (\ref{Faddeevfvdiscret}) may now be treated as secular equations with $ T^{(L)}_{\beta, j}$ as coefficients of the expansion basis, and thus the quantization condition for multiparticles in a finite volume is given by:
\begin{equation}
\det  \left [ \delta_{\alpha, \beta}  \delta_{i,j}      -  ( v_\alpha G_L   )_{i, j}  \right ] =0.   \label{multibodyqcdiscret}
\end{equation}
 Now, the interaction potentials can be parameterized and used as inputs for the quantization condition. The potentials may be extracted by fitting lattice data and infinite volume dynamics, such as the scattering amplitudes, can then be computed in separate steps.
 For multiparticle dynamics, dealing with potentials directly is typically much easier than handling multiparticle scattering amplitudes. We note that the HAL QCD collaboration method is also an interesting approach for extracting interaction potentials from lattice data \cite{Doi:2011gq}. A four-particle example is presented in Appendix to further illustrate the variational approach for multiparticle dynamics.

{\bf  Summary.}-- In this study, we proposed a general framework for finding multiparticle propagating solutions on a torus. We derived multiparticle secular equations in a finite volume based on the variational approach combined with the Faddeev method, where periodic boundary conditions are imposed by the finite volume Faddeev type or Lippmann-Schwinger type integral equations. Both the multiparticle quantization condition and finite volume wave function solutions may be obtained from the secular equations. Two versions of the multiparticle secular equations are presented in Eq. (\ref{multibodyseq}) and Eq. (\ref{Faddeevfvdiscret}). The first version of the secular equations given in Eq. (\ref{multibodyseq}) and the corresponding quantization condition in Eq. (\ref{multibodyqc1}) or Eq. (\ref{multibodyqc2}) are derived based on the assumption that the finite volume variational trial wave function is given by the linear superposition of all possible infinite volume solutions: $\Phi = \sum_J c_J \Psi_J$. Hence, version one may reveal the close relationships between the propagating solutions on a torus and the solutions in an infinite volume. The second version of secular equations given in Eq. (\ref{Faddeevfvdiscret}) and the corresponding quantization condition in Eq. (\ref{multibodyqcdiscret}) may seem more obscure in terms of extracting infinite volume dynamical information for multiparticle interactions, but they may be more efficient and robust for specific tasks, such as lattice QCD data analysis.

 {\bf Acknowledgments.}--We acknowledge support from the Department of Physics and Engineering, California State University, Bakersfield, CA. We also acknowledge partial support by the National Science Foundation under Grant No. NSF PHY-1748958.

\appendix

\section{Example of four particles interaction on a torus}
As a specific example of multiparticle interactions in a finite volume, let us consider a four-particle system interacting via only pairwise interactions in a one-dimensional box. After removing the center of mass motion, the dynamics of the system in the rest frame for four particles is described by:
\begin{align}
 \phi   & (r_{14}, r_{24} ,r_{34})   =  \int_{ -\frac{L}{2}}^{\frac{L}{2}} d r'_{14} d r'_{24}  d r'_{34}  \nonumber \\
& \times G_L (r_{14} - r'_{14}, r_{24} -r'_{24} ,r_{34} - r'_{34} ) \nonumber \\
& \times  \sum_{ (  \alpha < \beta ) }^4 V_{( \alpha \beta )} (r'_{  \alpha \beta})   \phi (r'_{14}, r'_{24} ,r'_{34}), \label{4bLSwav}
\end{align}
where $\phi  $ is a wave function that describes the relative motion of the four-particle system, the independent relative coordinates are selected as $(r_{14}, r_{24} ,r_{34}) $,  and $V_{(\alpha \beta)} (r_{\alpha \beta})$ represents the pairwise interactions between the $\alpha$-th and $\beta$-th particles. The four-particle finite volume Green's function is given by:
\begin{equation}
G_L (r_{14} , r_{24}  ,r_{34}   )   = \frac{1}{L^3} \sum_{p'_1,p'_2,p'_3}  \frac{ e^{i  ( p'_1 r_{1 4}  + p'_2 r_{2  4}  + p'_3 r_{3 4}   ) }  }{E- \sum_{i=1}^4 \frac{ {p'_i}^2}{2m } }, 
\end{equation}
where $p'_{1,2,3}  \in \frac{2\pi}{L} n $, $n \in \mathbb{Z}$, and $p'_4 = - p'_1-p'_2-p'_3$. To simplify presentation, we assume that all particles are distinguishable but they carry the same mass: $m_1=m_2=m_3=m_4=m$. 

The six finite volume amplitudes may be introduced as:
\begin{align}
& T_{ ( \alpha \beta)}^{(L)} ( p_1 , p_2 , p_3  ) = -  \int_{ -\frac{L}{2}}^{\frac{L}{2}} d r_{14} d r_{24}  d r_{34}  \nonumber \\
&\times e^{- i  ( p_1 r_{1 4}  + p_2 r_{2  4}  + p_3 r_{3 4}   ) }   V_{( \alpha \beta )} (r_{ \alpha \beta})   \phi (r_{14}, r_{24} ,r_{34}),
\end{align}
where $(\alpha \beta ) = (14), (24), (34), (12), (13), (23)$. Thus, Eq. (\ref{4bLSwav}) is transformed into six coupled equations in terms of the $T_{ ( \alpha \beta)}^{(L)} $ amplitudes:
\begin{align}
& T_{ ( \alpha \beta)}^{(L)} ( p_1 , p_2 , p_3  )=   \sum_{ p'_1, p'_2, p'_3}   \delta_{p_\gamma, p'_\gamma }   \delta_{p_\delta, p'_\delta} \widetilde{V}_{(\alpha \beta)} ( q_{\alpha \beta}  - q'_{\alpha \beta} ) \nonumber \\
& \times \frac{1}{L} \frac{ 1 }{ E- \sum_{i=1}^4 \frac{ {p'_i}^2}{2m } } \sum_{ (\alpha' \beta')=1 }^4  T_{ ( \alpha' \beta' )}^{(L)} ( p'_1 , p'_2 , p'_3  ), \nonumber \\
& \quad \quad \quad \quad     \quad \quad \quad \quad     \quad \quad     \quad \quad  \quad \quad     \alpha \neq \beta \neq \gamma \neq \delta,
\end{align}
where $q_{\alpha \beta} = \frac{p_\alpha - p_\beta}{2}$ and $q'_{\alpha \beta} = \frac{p'_\alpha - p'_\beta}{2}$ are the relative momenta between the $\alpha$-th and $\beta$-th particles, and the momenta of the four particles are constrained by the total momentum conservation $p_\alpha + p_\beta + p_\gamma + p_\delta =0 $. $\widetilde{V}_{(\alpha \beta)} $ represents the Fourier transform of potential $V_{(\alpha \beta)} $: $\widetilde{V}_{(\alpha \beta)} ( q_{\alpha \beta}  )  = \int_{ -\frac{L}{2}}^{\frac{L}{2}} d r_{\alpha \beta}  e^{ - i  q_{\alpha \beta} } V_{(\alpha \beta)} ( r_{\alpha \beta}  )  $. Thus, the quantization condition is given by:
\begin{align}
& \det \bigg [    \delta_{(\alpha \beta), (\alpha' \beta')} \prod_{i=1}^3 \delta_{p_i, p'_i}   -   \frac{ 1}{L} \frac{  \delta_{p_\gamma, p'_\gamma }   \delta_{p_\delta, p'_\delta} \widetilde{V} ( q_{\alpha \beta}  - q'_{\alpha \beta} ) }{ E- \sum_{i=1}^4 \frac{ {p'_i}^2}{2m } } \bigg ]  \nonumber \\
&    =0 ,  \quad \quad \quad \quad \quad \quad \quad \quad       \quad \quad \quad \quad       \alpha \neq \beta \neq \gamma \neq \delta.
\end{align}


\begin{thebibliography}{99}








\bibitem{Kambor:1995yc}
J.~Kambor, C.~Wiesendanger, and D.~Wyler, 
Nucl.\ Phys.\ {\bf B465}, 215 (1996).

\bibitem{Anisovich:1996tx}
A.~V.~Anisovich and H.~Leutwyler, 
Phys.\ Lett.\ {\bf B375}, 335 (1996).

 
\bibitem{Schneider:2010hs}
S.~P.~Schneider, B.~Kubis, and C.~Ditsche, 
JHEP {\bf 1102}, 028 (2011).



\bibitem{Guo:2015zqa}
P.~Guo, Igor~V.~Danilkin, D.~Schott, C.~Fern\'andez-Ram\'{\i}rez, V.~Mathieu, and A.~P.~Szczepaniak, Phys.\ Rev.\ {\bf D92}, 054016 (2015).


\bibitem{Guo:2016wsi}
P.~Guo, Igor~V.~Danilkin, C.~Fern\'andez-Ram\'{\i}rez, V.~Mathieu, and A.~P.~Szczepaniak, 
Phys.\ Lett.\ {\bf B771}, 497 (2017).


\bibitem{Colangelo:2016jmc}
 G.~Colangelo, S.~Lanz, H.~Leutwyler, and E.~Passemar, 
Phys.\ Rev.\ Lett.\ {\bf 118}, 022001 (2017).


 



\bibitem{Efimov:1970zz} 
V.~Efimov,
Phys.\ Lett.\ {\bf B 33}, 563 (1970).


\bibitem{Zhukov:1993aw} 
M.~V.~Zhukov, B.~V.~Danilin, D.~V.~Fedorov, J.~M.~Bang, and I.~J.~Thompson, 
Phys.\ Rept.\ {\bf 231}, 151 (1993).


 


 


  \bibitem{Luscher:1990ux}
M.~L\"uscher,
  Nucl.\ Phys.\ B {\bf 354}, 531 (1991).



  \bibitem{Rummukainen:1995vs}
K.~Rummukainen and S.~Gottlieb,
  Nucl.\ Phys.\ B {\bf 450}, 397 (1995).

             
  \bibitem{Christ:2005gi}
	N.~H.~Christ, C.~Kim, and T.~Yamazaki,
  Phys.\ Rev.\ D {\bf 72}, 114506 (2005).
            
  \bibitem{Bernard:2007cm}
V.~Bernard, Ulf-G.~Mei{\ss}ner, and A.~Rusetsky,
  Nucl.\ Phys.\ B {\bf 788}, 1 (2008).
  
  \bibitem{Bernard:2008ax}
V.~Bernard, M.~Lage, Ulf-G.~Mei{\ss}ner, and A.~Rusetsky,
  JHEP {\bf 0808}, 024 (2008).
  
   
  \bibitem{He:2005ey}
S.~He, X.~Feng, and C.~Liu,
  JHEP {\bf 0507}, 011 (2005).



  \bibitem{Lage:2009zv}
M.~Lage, Ulf-G.~Mei{\ss}ner, and A.~Rusetsky,
Phys.\ Lett.\ B {\bf 681}, 439 (2009).




  \bibitem{Doring:2011vk}
M.~D\"oring, Ulf-G.~Mei{\ss}ner, E.~Oset and A.~Rusetsky,
Eur.\ Phys.\ J.\ A {\bf 47}, 139 (2011).


    

\bibitem{Briceno:2012yi} 
  R.~A.~Briceno and Z.~Davoudi,
  Phys.\ Rev.\ D {\bf 88}, 094507 (2013).
  


\bibitem{Hansen:2012tf} 
  M.~T.~Hansen and S.~R.~Sharpe,
  Phys.\ Rev.\ D {\bf 86}, 016007 (2012).
  



\bibitem{Guo:2012hv} 
  P.~Guo, J.~Dudek, R.~Edwards, and A.~P.~Szczepaniak,
  Phys.\ Rev.\ D {\bf 88}, 014501 (2013).
  


  \bibitem{Guo:2013vsa}
  P.~Guo, 
  Phys.\ Rev.\ D {\bf 88}, 014507 (2013).





\bibitem{Kim:2005gf} 
  C.~H.~Kim, C.~T.~Sachrajda, and S.~R.~Sharpe,
  Nucl.\ Phys.\ B {\bf 727}, 218 (2005).
  



\bibitem{Briceno:2014uqa} 
  R.~A.~Briceno, M.~T.~Hansen, and A.~Walker-Loud,
  Phys.\ Rev.\ D {\bf 91}, 034501 (2015).
  


\bibitem{Romero-Lopez:2018zyy} 
  F.~Romero-L\"opez, A.~Rusetsky, and C.~Urbach,
  Phys.\ Rev.\ D {\bf 98}, 014503 (2018).
  







  \bibitem{Kreuzer:2008bi}
	S.~Kreuzer and H.-W.~Hammer,
	Phys.\ Lett.\ B {\bf 673}, 260 (2009).
	
	

  \bibitem{Kreuzer:2009jp}
	S.~Kreuzer and H.-W.~Hammer,
	Eur.\ Phys.\ J.\ A {\bf 43}, 229 (2010).
	
	

  \bibitem{Kreuzer:2012sr}
	S.~Kreuzer and H.-W.~Hammer,
	Eur.\ Phys.\ J.\ A {\bf 48}, 93 (2012).
		


  \bibitem{Polejaeva:2012ut}
	K.~Polejaeva and A.~Rusetsky,
	Eur.\ Phys.\ J.\ A {\bf 48}, 67 (2012).
		



  \bibitem{Briceno:2012rv}
	 R.~A.~Briceno and Z.~Davoudi,
	Phys.\ Rev.\ D {\bf 87}, 094507 (2013).





  \bibitem{Hansen:2014eka}
	 M.~T.~Hansen and S.~R.~Sharpe,
	Phys.\ Rev.\ D {\bf 90}, 116003 (2014).





  \bibitem{Hansen:2015zga}
	 M.~T.~Hansen and S.~R.~Sharpe,
	Phys.\ Rev.\ D {\bf 92}, 114509 (2015).



  \bibitem{Hansen:2016fzj}
	 M.~T.~Hansen and S.~R.~Sharpe,
	Phys.\ Rev.\ D {\bf 93}, 096006 (2016).






\bibitem{Hammer:2017uqm}
H.~-W.~Hammer, J.~-Y.~Pang, and A.~Rusetsky, JHEP {\bf 1709}, 109 (2017).



\bibitem{Hammer:2017kms}
H.~-W.~Hammer, J.~-Y.~Pang, and A.~Rusetsky, 
JHEP {\bf 1710}, 115 (2017).



  \bibitem{Meissner:2014dea}
Ulf-G.~Mei{\ss}ner, G. Rios, and A.~Rusetsky,
 Phys.\ Rev.\ Lett.\ {\bf 114}, 091602 (2015). Erratum: Phys.\ Rev.\ Lett. {\bf 117}, 069902 (2016).





 \bibitem{Briceno:2017tce}
R.~A.~Briceno, M.~T.~Hansen, and S.~R.~Sharpe, 
Phys.\ Rev.\ {\bf D95}, 074510 (2017).



 \bibitem{Sharpe:2017jej}
S.~R.~Sharpe, 
Phys.\ Rev.\ {\bf D96}, 054515 (2017).


 \bibitem{Mai:2017bge}
M.~Mai and M.~D\"oring, 
Euro.\ Phys.\ J\ {\bf A53}, 240 (2017).
  




  \bibitem{Romero-Lopez:2018rcb}
F.~Romero-L\"opez, A.~Rusetsky, and C.~Urbach,
Euro.\ Phys.\ J\ {\bf C78}, 846 (2018).






\bibitem{Guo:2016fgl}
P.~Guo, 
Phys.\ Rev.\ {\bf D95}, 054508 (2017).




\bibitem{Guo:2017ism}
P.~Guo and V.~Gasparian, 
Phys.\ Lett.\ {\bf B774}, 441 (2017).





\bibitem{Guo:2017crd}
P.~Guo and V.~Gasparian, 
Phys.\ Rev.\ {\bf D97}, 014504 (2018).






  \bibitem{Guo:2018ibd}
P.~Guo, M.~D\"oring, and A.~P.~Szczepaniak,
  Phys.\ Rev.\ D {\bf 98}, 094502 (2018).
  
  


  \bibitem{Guo:2018xbv}
P.~Guo and T.~Morris,
Phys.\ Rev.\ {\bf D99}, 014501 (2019).
  



  \bibitem{Mai:2018djl}
M.~Mai and M.~D\"oring,
Phys.\ Rev.\ Lett.\ {\bf 122}, 062503 (2019).
  


  \bibitem{Briceno:2018aml}
R.~A.~Briceno, M.~T.~Hansen, and S.~R.~Sharpe, 
Phys.\ Rev.\ {\bf D99}, 014516 (2019).
  


  \bibitem{Romero-Lopez:2019qrt}
F.~Romero-L\"opez, S.~R.~Sharpe, T.~D.~Blanton, R.~A.~Briceno, and M.~T.~Hansen, 
JHEP {\bf D1910}, 007 (2019).
  


  \bibitem{Blanton:2019vdk}
T.~D.~Blanton, F.~Romero-L\"opez, and S.~R.~Sharpe,   
arXiv:1909.02973 [hep-lat]. 
  



  \bibitem{Doring:2018xxx}
M.~D\"oring, H.~-W.~Hammer, M.~Mai, J.~-Y.~Pang, A.~Rusetsky, and J.~Wu, 
Phys.\ Rev.\ {\bf D97}, 114508 (2018).
  


  \bibitem{Pang:2019dfe}
 J.~-Y.~Pang, J.~Wu, H.~-W.~Hammer, Ulf-G.~Mei{\ss}ner, and A.~Rusetsky,
Phys.\ Rev.\ {\bf D99}, 074513 (2019).
  






 

 \bibitem{Ziman:1964}
	J.~M.~Ziman,
  {\it Principles of the Theory of Solids}, 2nd Edition, Cambridge University Press, UK (1972), ISBN 978-1-1396-4407-5.






  \bibitem{Faddeev:1960su}
	L.~D.~Faddeev,
  Zh.\ Eksp. \ Teor.\ Fiz. \ {\bf 39}, 1459 (1960) [Sov. Phys.-JETP {\bf 12}, 1014 (1961)]. 



  \bibitem{Faddeev:1965}
	L.~D.~Faddeev,
  {\it Mathematical Aspects of the Three-Body Problem in the Quantum Scattering Theory}, Israel Program for Scientific Translation, Jerusalem, Israel (1965).





\bibitem{Gloeckle:1983}
W.~Gl\"ockle, 
 {\it The Quantum Mechanical Few-Body Problem}, Springer, Berlin, Germany (1983), ISBN 978-3-642-82083-0.



  \bibitem{Faddeev:1993}
	L.~D.~Faddeev and S.~P.~Merkuriev,
  {\it Quantum Scattering Theory for Several Particle Systems}, Springer, Netherlands (1993), ISBN 978-0-7923-2414-0.






\bibitem{Gloeckle:1995jg}
W.~Gl\"ockle, H.~Witala, D.~H\"uber, H.~Kamada, and J.~Golak, 
Phys.\ Rept.\ {\bf 274}, 107 (1996).





 


\bibitem{Doi:2011gq}
T.~Doi, {\it et al.} (HAL QCD Collaboration),
Prog.\ Theor.\ Phys.\ {\bf 127}, 723 (2012).




 
    
    
    
    
\end{thebibliography}
\end{document}